\documentclass[useAMS,usenatbib]{mn2e}
\usepackage{amssymb}
\usepackage{epsfig}
\usepackage{times}

\title[Orientation of Haloes around voids]
{The orientation of galaxy dark matter haloes around cosmic voids}

\author[R. Brunino et al.]{Riccardo Brunino$^{1}$, Ignacio Trujillo$^{1}$,
 Frazer R. Pearce$^{1}$ and Peter A. Thomas$^{2}$\\
 $^{1}$School of Physics and Astronomy, University of Nottingham,
   University Park, Nottingham NG7 2RD, UK\\
   $^{2}$ Astronomy Centre, University of Sussex, Falmer, Brighton, BN1 9QH, UK}

\begin{document}

\date{}

\pagerange{\pageref{firstpage}--\pageref{lastpage}} \pubyear{2006}

\maketitle

\label{firstpage}

\begin{abstract}

Using the Millennium N--body simulation we explore how the shape and  angular
momentum of galaxy dark matter haloes  surrounding the largest cosmological
voids are oriented. We find that the major and intermediate axes of the haloes
tend to lie parallel to the surface of the voids, whereas the minor axis points
preferentially in the radial direction. We have quantified the strength of these
alignments at different radial distances from the void centres. The effect of
these orientations is still detected at distances as large as 2.2 R$_{void}$
from the void centre. Taking a subsample of haloes expected to contain 
disc--dominated galaxies at their centres we detect, at the 99.9\% confidence
level,  a signal that the angular momentum of those haloes tends to lie parallel
to the surface of the voids. Contrary to the alignments of the inertia axes,
this signal is only detected in shells at the void surface (1$<$R$<$1.07
R$_{void}$) and disappears at larger distances. This signal, together with the
similar alignment observed using real spiral galaxies (Trujillo, Carretero \&
Patiri 2006), strongly supports the prediction of the Tidal Torque theory  that
both dark matter haloes and baryonic  matter have acquired, conjointly, their
angular momentum before the moment of turnaround.

\end{abstract}


\section{Introduction}

The angular momentum  of a galaxy  plays a central role in determining its
evolution and final type. However, understanding the origin and properties of
the galactic angular momentum  has been one of the key problems in astrophysics
in the last five decades. The current 'standard' theory for the origin of the
angular momentum, within the cosmological framework of  hierarchical structure
formation, is the tidal--torque theory (hereafter TTT). This theory is based on
early ideas from Hoyle (1951) that suggested that the angular momentum of a
galaxy arises from the tidal field of neighbouring galaxies. This idea was
further developed and quantified by Peebles (1969), Doroshkevich (1970) and
White (1984).

The TTT suggests that most of the angular momentum is gained gradually by the
protohaloes during the linear regime of the growth of density fluctuations, due
to  tidal torques from neighbouring fluctuations. Angular momentum grows
linearly with time at this early epoch and saturates when
the halo decouples from the expanding background at the moment of turnaround.
It is assumed that during this phase the baryonic component follows the dark
matter distribution and consequently gains a similar specific angular momentum
to that of the halo. A subsequent large collapse factor of the baryonic matter
to the centre of the halo will explain the centrifugally supported nature of
the galactic discs (Fall \& Efstathiou 1980).

To first order, the angular momentum of the haloes results from the misalignment
between the principal axes of the inertia momentum tensor (I$_{ij}$) of the
matter being torqued and the principal axes of the shear or tidal tensor
(T$_{ij}$=-$\partial^2\phi/\partial x_i\partial x_j$) generated by neighbouring
density fluctuations. The leading term of the torque is given by
L$_i$$\propto$T$_{jk}$(I$_{jj}$-I$_{kk}$), where $i$, $j$, and $k$ are cyclic
permutations of 1, 2 and 3. Several aspects of this picture have been confirmed
by a number of studies of the angular momentum properties of dark matter haloes
both analytically  (Heavens \& Peacock 1988; Catelan \& Theuns 1996) and in
N--body simulations (Barnes \& Efstathiou 1987; Sugerman, Summers \&
Kamionkowski 2000; Lee \& Pen 2000; Porciani, Dekel \& Hoffman 2002a,b).

If $\bf{I}$ and $\bf{T}$ were uncorrelated (as frequently has been assumed under
the argument that the former is local and the latter is dominated by the
distribution of the large--scale structure) the direction of the angular
momentum in the linear regime should be aligned with the intermediate axis of
$\bf{I}$ (the direction that maximizes the difference between I$_{jj}$ and
I$_{kk}$). However, Porciani et al. (2002b) have found in their simulations that
there is a strong correlation between the $\bf{I}$ and $\bf{T}$ tensors, in the
sense that their minor, major and intermediate principal axes tend to be aligned
at the protohalo stage. The strong correlation between $\bf{I}$ and $\bf{T}$
should produce an angular momentum vector of the haloes that is perpendicular to
the minor axis of the sheet  they are embedded in (i.e.  perpendicular to the 
direction of the maximum compression of the large--scale structure at that
point). However, this last correlation, at least at redshift zero, is expected
to be very weak (or totally erased) by non--linear effects (such as exchange of
angular momentum between haloes) at late times.

From the observational point of view, Trujillo, Carretero \& Patiri
(2006) have found that the rotation axes of spiral galaxies located on the
shells of the largest cosmic voids lie preferentially on the void surface (in
agreement with the expectation for the angular momentum of haloes given in the
previous picture). The observational relation could be explained then as a
consequence of the spin of the baryonic matter still retaining memory of the
angular momentum properties of the haloes at the moment of  turnaround
(Navarro, Abadi \& Steinmetz 2004). It is key, consequently, to explore whether
the signal is also found in the haloes of cosmological N--body simulations when
we mimic the observational technique. If so, this will strongly support our
current understanding of how haloes and baryonic matter have acquired,
conjointly, their angular momentum.

Together with the orientation of the angular momentum, the alignment of the
shape of the galaxy dark matter haloes (M$<$10$^{13}h^{-1}$M$_{\sun}$) with
their surrounding large--scale structure can have important observational
consequences. In fact, the shapes of dark matter haloes can affect the coherence
of tidal streams (Sackett 1999), can be related to galactic warps (Ostriker \&
Binney 1989; Debattista \& Sellwood 1999; L\'opez--Corredoira et al. 2002) or
can affect the distribution of the orbits of satellite galaxies (Holmberg 1969;
Zaritsky et al. 1997; Sales \& Lambas 2004; Agustsson \& Brainerd 2006; Yang et
al. 2006). In addition, infall of material into the haloes is not    isotropic
but  is expected to be through the filaments where the haloes are embedded.
Consequently, the orientation of the dark matter haloes within these structures
can affect the characteristics of the galaxy properties previously mentioned.

The alignment of massive (group and cluster) haloes
(M$>$10$^{13}h^{-1}$M$_{\sun}$) with their surrounding large--scale structure has
been explored in detail (Splinter et al. 1997; Onuora \& Thomas 2000; Faltenbacher
et al. 2002; Kasun \& Evrard 2005; Hopkins et al. 2005; Basilakos et al. 2006).
These works indicate that the major axes of neighbouring galaxy clusters are
aligned. A result that is in agreement with the "Binggeli (1982) effect". The
origin of these alignments is still under debate and could be associated to infall
of material (van Haarlem \& van de Weygaert 1993) and/or tidal fields (Bond et al.
1996). Due to the lack of resolution in previous simulations, it is only now that
an exploration of the alignment of the galaxy dark matter haloes has become 
possible (Bailin \& Steinmetz 2005, hereafter BS05; Altay, Colberg \& Croft 2006,
hereafter ACC06; Patiri et al. 2006b, hereafter PA06; Arag\'on--Calvo et al.
2006). These works suggest that the major axis of the haloes lies along the
filaments. The quantification of the strength of these alignments is key to
studies of strong and weak lensing where the intrinsic distribution and alignment
of galaxy shapes plays an important role in interpreting the signal (see e.g.
Heavens, Refregier \& Heymans 2000; Croft \& Metzler 2000; Heymans et al. 2006).

The aim of this paper is to characterise the alignment of both the shape and
angular momentum of galaxy dark matter haloes with their surrounding
large--scale structure to an unprecedent statistical level using the Millennium
simulation (Springel et al. 2005).  In particular, we will focus our attention
on haloes surrounding the largest cosmological voids. Contrary to filaments
(which are strongly affected by redshift--space distortion), large cosmological
voids are a feature easy to characterise from the observational point of view
(Trujillo et al. 2006). In addition, another important advantage of the void
scheme is that (because of the radial growth of the voids) the vector joining
the centre of the void with the galaxy (halo) is a good approximation to the
direction of the maximum compression of the large--scale structure at that point.

This naturally generates a framework for exploring the alignments of the shape
and angular momentum of the haloes with their surrounding matter distribution.
Consequently, our work mimics the observational framework to provide an easy
interpretation of current and future observations. The large volume sampled by
the Millennium simulation combined with the excellent spatial resolution allows
us to conduct this project. This  unprecendent statistical power is absolutely
crucial if we want to explore signals expected to be very weak like the
alignment of  halo angular momentum and the large--scale structure.

This paper is structured as follows. Section 2 provides a description of the
Millennium simulation itself, the void and halo samples used, and details of how
the shapes and spins were measured. Section 3 describes our results and we
discuss our findings in Section 4.

\section[]{Numerical simulation}

\subsection[]{N--body simulation}

The main simulation we have used for this study is the {\it Millennium
Simulation} of Springel et al (2005). This employs $2160^3$ dark matter
particles each of mass $8.6\times 10^8 h^{-1}{\rm M}_\odot$
within a comoving box of side $500h^{-1}{\rm Mpc}$. This simulation
was performed in a $\Lambda$CDM universe with cosmological parameters:
$\Omega_\Lambda = 0.75, \Omega_M = 0.25, \Omega_b = 0.045, h = 0.73, n=1$,
and $\sigma_8 = 0.9$, where the Hubble constant is characterised as
$100h {\rm km s^{-1} Mpc^{-1}}.$ These cosmological parameters are
consistent with recent combined analyses from {\it WMAP} data
(Spergel et al 2003) and the 2dF galaxy redshift survey (Colless et al
2001), although the value for $\sigma_8$ is a little higher than would
perhaps have been desirable in retrospect. The spatial resolution is
$5h^{-1}{\rm kpc}$ everywhere inside the simulation volume.

\subsection[]{Void and halo samples}

To explore the effect of the large--scale structure on the orientation of the
dark matter haloes, we use a similar technique to the one adopted in analysing
the orientation of disc galaxies  in real observations (Trujillo et al. 2006).
To this aim we need, first, to find and characterise the radius of the large
voids in the simulation. To speed up the process of finding the voids, we used
as an initial guess for the position of the void centres the position of the most
underdense particles in the box. To do this, for every dark matter particle in
the Millennium Simulation we have estimated the local density by using a
standard (Monaghan \& Lattanzio 1985; Hernquist \& Katz 1989) smoothing kernel
averaging over 32 neighbours. We produced a list of the most underdense
particles and then radially sorted surrounding particles in distance from these
points. In this way the radius of a volume underdense by a factor of 10, centred
on the most underdense particle, was calculated. All particles with densities
less than 0.035 of the cosmic mean and further than $2h^{-1}{\rm Mpc}$ from an
even less dense particle were tried as prospective centres. After cleaning this
catalogue  by removing  those positions
that lay within a larger void and limiting the size to be larger than $5h^{-1}{\rm Mpc}$ 
in radius, this technique produced 3024 potential void centre candidates.               
We estimate that this catalogue of potential void centres is
greater than $99\%$ complete, given that we tried over 200,000 random centres
but only found 2 additional voids in the last 50,000 (i.e. voids with central
densities close to our limit of 0.035).

Once a list of initial positions for searching for voids in the simulation had
been created, we used these positions to search for the maximum spheres that
are empty of haloes with mass larger than a given value. To conduct this
search we follow  an algorithm  based on a modification of the one
presented in Patiri et al. (2006a). The haloes were identified using a minimal
spanning tree that links together particles with density exceeding 900 times
the background density (Thomas et al. 1998). This is to focus on the core
properties. In our case, we have used all haloes with masses larger than
8.6$\times$10$^{11} h^{-1}$M$_{\sun}$ (i.e. those haloes with more than 1000
particles). Using this mass cut we are selecting haloes that contain galaxies
with stellar masses similar to the ones used in the observational data. To
characterise the final position of the void centres and their radii, we
populate a sphere of radius R=5 h$^{-1}$ Mpc, centered on each initial
position, with 2000 random points. For every point in this sphere, we estimate
the position of the closest 4 haloes lying in geometrically "independent"
octants. Then we built the sphere 
defined by these 4 haloes. This is repeated
for all the 2000 random points. As a characterization (position and radius) of
the void we choose the biggest empty spherical region generated in the
previous step.  It is important to stress that the position of the void
defined in this way normally does not match the position of the initial guess.

To match the observations we select only those voids whose radius (as measured
by the largest sphere that is empty of haloes greater than a given mass) is
larger than 10 $h^{-1}$ Mpc. This cut produces 2932 voids in our box with a
median radius of 14 $h^{-1}$ Mpc. To explore the alignments of the haloes we
have concentrated only on haloes with masses 8.6$\times$10$^{11}
h^{-1}$M$_{\sun}$ $<$M$<$ 8.6$\times$10$^{12} h^{-1}$M$_{\sun}$(i.e. we restrict
the sample to galaxy--sized haloes) and located within shells beyond the surface
of the voids.

In addition, we have also created a subsample of haloes which contain a
disc--dominated galaxy at their centre. These haloes are expected to have had a
relatively 'quiet' life and have suffered smaller non--linear effects after 
turnaround such that their spin properties should remain matched to those of the
baryonic matter. To select these haloes we have used the semianalytic galaxy
catalogue of Croton et al. (2006). We have created the subsample  by selecting the
brightest dominant galaxy within 200 $h^{-1}$ kpc of our halo centre. We then select
those haloes where the semianalytic code indicates that there is a galaxy brighter
than M$_{K}$$<$-23 mag that has a bulge--to--total (B/T) ratio 0$<$B/T$<$0.4 (i.e. a
Milky--Way like disc galaxy).

\subsection[]{Halo shapes and spins}

To characterise the orientation of the haloes we have determined the orientation
of their principal axes and their angular momentum vectors. The principal axes
of the halo mass distribution are measured by diagonalising the inertia tensor,

\begin{equation}
I_{ij}=\frac{1}{N_p}\sum_{k=1}^{N_p}m_kx_{k,i}x_{k,j}
\end{equation}

where the sum is over all the particles in the halo N$_p$, and the coordinates are
defined with respect to the centre of  mass of the halo of mass M$_h$. The
resulting eigenvalues M$_h$a$^2$/5, M$_h$b$^2$/5, and M$_h$c$^2$/5 are sorted by
size,  in descending order. The eigenvectors give the directions of the principal
axes.

The angular momentum vector of the halo is given by:

\begin{equation}
\bold{L}=\sum_{k}m_{k}\bold{r_k}\times\bold{v_k}
\end{equation}

where the sum is again over all the particles in the halo and $\bold{r_k}$ and
$\bold{v_k}$ are the position and velocity of each particle relative to the
centre of mass of the halo.

The uncertainty in the position of the inertia axes and of the angular momentum has
been evaluated by comparing these quantities in the same haloes in two different
runs that differ by a factor of 20 in resolution.  We found that we can measure the
orientation of the angular momentum vector with an uncertainty (as provided by the
full width half maximum of the ditribution) of $\sim$14$^\circ$ when the number of
particles in the (low resolution) halo is larger than 1000. The uncertainties in
the inertia axes are 13$^\circ$ for the semimajor and semiminor axes and 20$^\circ$
for the intermediate axis. 

Once the angular momentum and the inertia vectors are evaluated, we estimate the
cosine of the angle between those vectors and the vector joining the centre of the
void with the centre of the halo $\bold{R}$:

\begin{equation}
\mu=\cos\theta=\big(\frac{\bold{R} \cdot \bold{V}}{|\bold{R}||\bold{V}|}\big)
\end{equation}

\section{Results}

\begin{figure*}

\epsfig{file=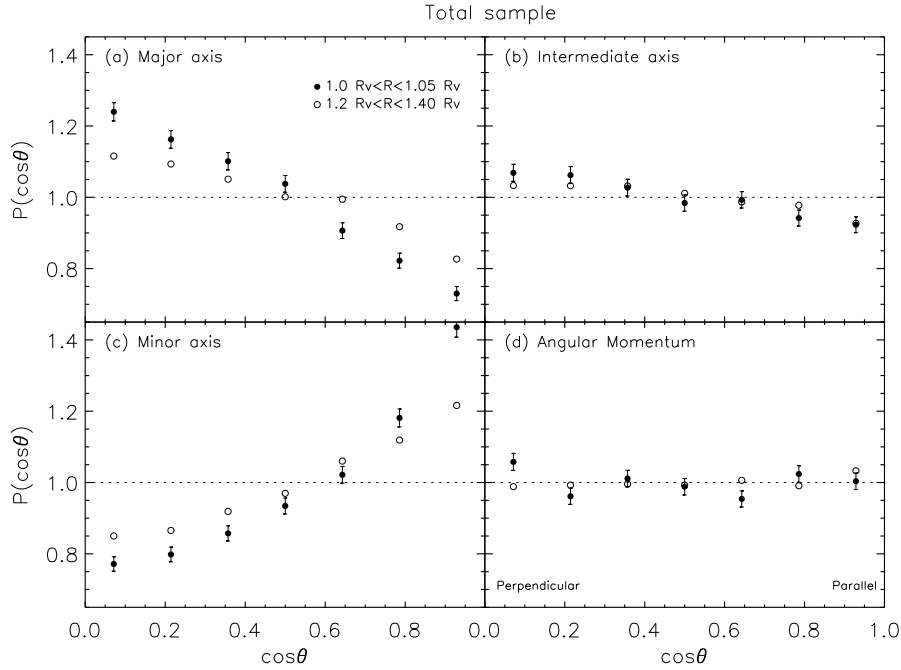,width=0.75\textwidth} 

\caption{Probability density distribution of the cosine of the angles $\theta$
between the inertia axes (and angular momentum) and the vector joining the
centre of the voids with the halo centres. The error bars on each bin is the
Poissonian error and (to avoid confussion) are only plotted for the innermost
shell bins.  The null hypothesis (i.e. a sine distribution) is represented by
the dashed line.}

\label{all}

\end{figure*}

Fig. \ref{all} shows the probability density distribution  P($\cos\theta$) of the
cosine of the angles  between the inertia axes (and the angular momentum) and the
vector joining the centre of the void to the centre of the halo. We show the
results for two different shells: the shell located at the surface of the void with
a width of 5\% of the  radius (i.e. 1 R$_{void}$$<$R$<$1.05 R$_{void}$), and a
shell located well beyond the surface of the void at 1.2 R$_{void}$$<$R$<$1.4
R$_{void}$. We do this to highlight  the effect of moving farther away from the
void surface. The dashed line in this figure corresponds to the probability
distribution of randomly distributed angles (i.e. P($\cos\theta$)=1).

To the best of our knowledge there is no theoretical prediction for the
probability density distribution of the angles  between the inertia axes and the
vector joining the centre of the void to the centre of the halo. For this
reason, we have used the following simple analytical expression motivated by
the planar symmetry of the problem (Betancort--Rijo \& Trujillo 2007; in
preparation) to quantify the strength of the signal:

\begin{equation} P(\mu)d\mu\propto\frac{pd\mu}{(1+(p^2-1)\mu)^{3/2}} \end{equation}

If p=1, we obtain the null hypothesis (i.e. P($\cos\theta$)=1). Values of p$>$1
imply that the inertial axis tends to be aligned with the surface of the shell, whereas
p$<$1 implies that the axis tends to be perpendicular to the void surface. This
simple analytical expression produces  good fits to the observed distribution with
reduced $\tilde\chi^2$$\lesssim$1 in most of the cases. The results of our fits are
summarized in Table \ref{data}.

We find significant alignments of the inertia axes within the
shells of the voids. The major axes of the dark matter haloes tends to lie
parallel to the surface of the voids (i.e. there is an excess of haloes with
large values of $\theta$).  For the minor axis the alignment is  contrary to
the major axis, there is an excess of haloes with the minor axis oriented in
the radial direction of the voids. The intermediate axis also tends to lie
parallel to the surface of the voids, although the signal is not as strong as
in the case of the major axis. As expected, the signal declines in all the
cases as  the distance from the centre of the voids is increased, although this
decline is  slow: we still detect a weak signal of alignments of
the major and minor axes at distances as large as $\sim$2.2 R$_{void}$ from the
void centre. The distribution of angular momentum vectors, however, is
compatible with a random orientation.

To test the reliability of our results we have repeated our analysis locating
the centres of the voids at random positions within the whole volume of the
simulation. As expected, we recovered for all the cases a signal which is
compatible with the null hypothesis. To run this test we have used exactly the
same number of random centres as the number of voids we have. The number of
haloes and their distances to the centres of these 'fake' voids are similar in this
control experiment to the real case.

\begin{figure*}

\epsfig{file=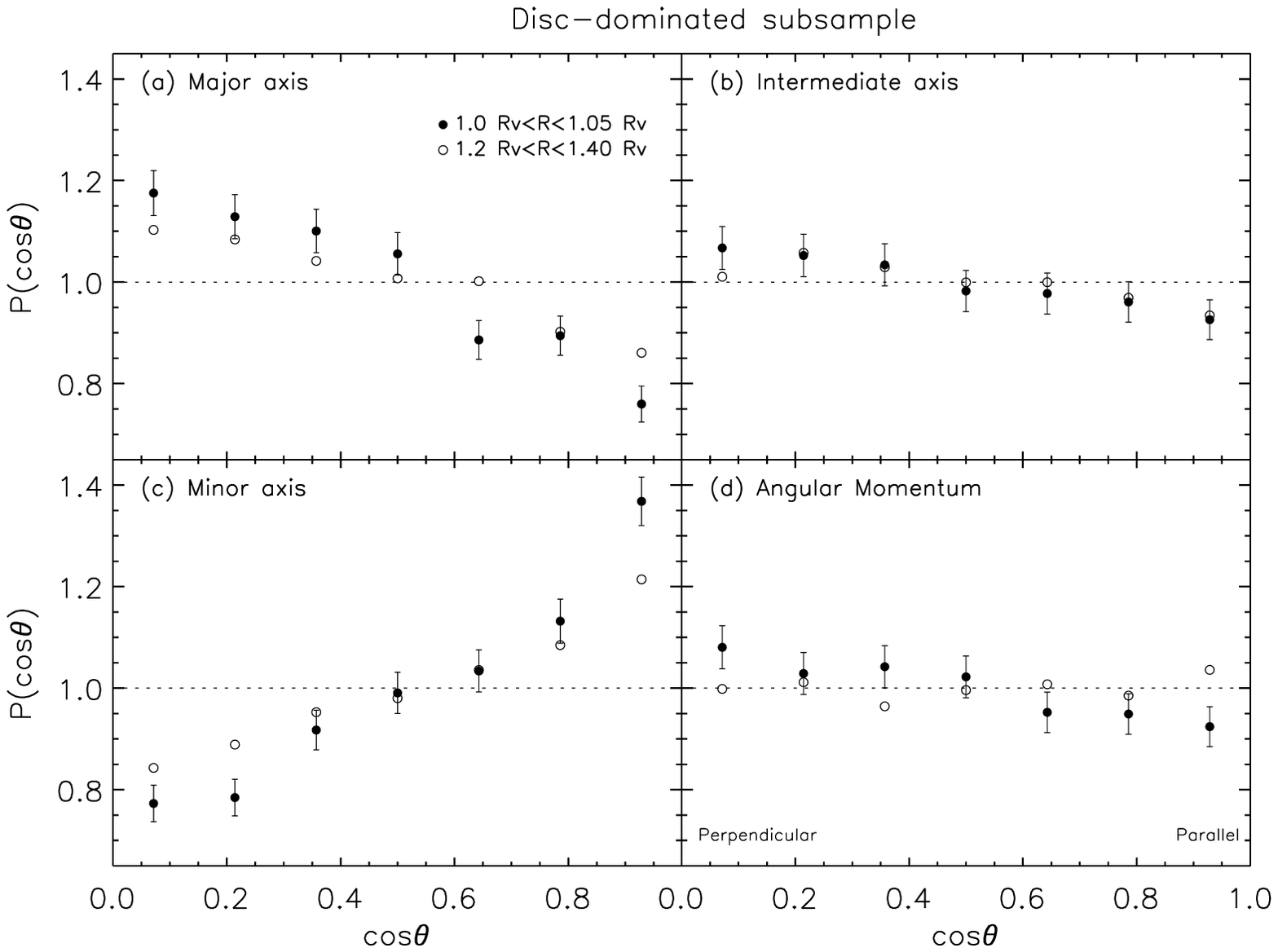,width=0.75\textwidth}

\caption{Same than in Fig. \ref{all} but this time using only haloes which
contain a disc--dominated galaxy at their centres. Note that using this halo
subsample the angular momentum vector tends to lie  parallel to the surface of
the void.} 

\label{discs}

\end{figure*}

As stated in the Introduction, any signal of alignment in the angular momentum
of the haloes is expected to be erased  after the turnaround by non-linear
effects. For this reason, if this signal is still present nowadays  it should
be found in haloes which have had the 'quietest' lives since  turnaround. To
explore this, we have repeated the same analysis as before but this time using 
the subsample of haloes with a disc-dominated galaxy at their centre. The
results are shown in Fig. \ref{discs} and the strength of the signal is
quantified in Table \ref{data}.

Contrary to the result obtained using the full sample of haloes, the angular
momentum vectors of the haloes with a disc--dominated galaxy at their centre
 tend to lie parallel to
the surface of the void. To test the reliability of this signal we have run
different statistical tests: the departure of the average (Avni \& Bahcall
1980) of $\cos \theta$ from 0.5 (i.e. the expected value in the null hypothesis
case) and the Kolmogorov--Smirnov (K-S) test. Both tests reject the null
hypothesis at the 99.8\% level. Our results are robust to changes in the ratio
B/T ranges selected (i.e. we still find a significant signal for the alignment
of the angular momentum including haloes with galaxies with B/T$<$0.6). In
addition, we have also checked that including  those haloes contained in our
list of voids with R$<$10 $h^{-1}$ Mpc do not alter (within the error bars) our
results. This is as expected beacause almost all of our initial void centres 
produce voids larger than 10 h$^{-1}$ Mpc in radius.

 It is worth noting that the maximum signal is found when we select a shell of
width 1 R$_{void}$$<$R$<$1.07 R$_{void}$. For this case, the null hypothesis is
rejected at 99.9\%. On the other hand, the inertia axes show the same trends in
this subsample as in the previous case using all the haloes.

To characterize the strength of the alignment of the angular momentum we have
followed two approachs. We have used (as before) Eq. (4) and, secondly, we    
have  compared our result with the theoretical prediction for this
quantity from Lee (2004) within the framework of the TTT. The strength of the
intrinsic galaxy alignment of the galaxies with local shear at the present
epoch is expressed as the following quadratic relation (Lee \& Pen 2002):

\begin{equation}
<L_iL_j>=\frac{1+c}{3}\delta_{ij}-c\widehat{T}_{ik}\widehat{T}_{kj},
\end{equation} 

where $\bold{L}$ is the halo angular momentum (spin) vector and
$\bold{\widehat{T}}$ is the rescaled traceless shear tensor $\bold{T}$  defined
as $\widehat{T}_{ij}$=$\widetilde{T}_{ij}/|\bold{\widetilde{T}}|$ with
$\widetilde{T}_{ij}$$\equiv$$T_{ij}-Tr(\bold{T})\delta_{ij}/3$, and c is a
correlation parameter introduced to quantify the strength of the intrinsic
shear-spin alignment in the range of [0,1]. To estimate c we have used the
analytical approximation suggested in Lee, Kang \& Jing (2005):

\begin{equation}
P(\mu)=\big( 1-\frac{3c}{4} \big)+\frac{9c}{8}(1-\mu^2)
\end{equation}

The values of the parameter c we
obtain in the different shells are summarized in Table \ref{data}. When c=1 the
strength of the galaxy alignment with the large-scale distribution is maximum,
whereas c=0 implies galaxies are oriented randomly. In the inner shell, we
measure in this case c=0.151$\pm$0.046. It is worth noting that the value of c
measured in this work is a lower limit of the true value because it has been
evaluated without any attempt to correct for the smoothing produced by the
uncertainty in measuring the angular momentum vector of the haloes.
Consequently, the strength (and statistical significance) of the observed
alignment should be higher. To quantify this effect, we have re--estimated c
again but this time using the theoretical prediction convolved with our error
function in measuring the angular momentum. After fitting the convolved
function to the data,  we obtain c=0.158$\pm$0.045.

\begin{table*}

 \centering


  \caption{The strength of the alignments on the different shells. To measure the alignment of the inertia
  axes we have used the parameter p (p=1 in the null case), and c (Lee 2004) in the case of the angular
  momentum (c=0 in the null case).}

  \begin{tabular}{rrrrrrr}

  \hline
Shell & Maj. A. & Int. A. & Min. A.  &  A. M. & A. M. & Number of \\
  R$_{void}$ Units   &  p  &  p	& p & p & c & Haloes\\
 \hline
& & & Total Sample   &\\
\hline
1.00 $<$R$<$1.05  & 1.218$\pm$0.012 & 1.061$\pm$0.011 & 0.778$\pm$0.008 & 1.007$\pm$0.010&  0.004$\pm$0.026 & 13128 \\
1.05 $<$R$<$1.10  & 1.165$\pm$0.014 & 1.060$\pm$0.013 & 0.800$\pm$0.010 & 0.994$\pm$0.012& -0.014$\pm$0.032 & 8644  \\
1.10 $<$R$<$1.20  & 1.150$\pm$0.009 & 1.057$\pm$0.008 & 0.820$\pm$0.007 & 0.992$\pm$0.008& -0.020$\pm$0.020 & 21567 \\
1.20 $<$R$<$1.40  & 1.111$\pm$0.005 & 1.040$\pm$0.005 & 0.863$\pm$0.004 & 0.989$\pm$0.005& -0.033$\pm$0.012 & 60908 \\
1.40 $<$R$<$1.80  & 1.070$\pm$0.003 & 1.021$\pm$0.003 & 0.914$\pm$0.002 & 0.989$\pm$0.003& -0.027$\pm$0.007 & 200201  \\
1.80 $<$R$<$2.60  & 1.030$\pm$0.001 & 1.011$\pm$0.001 & 0.959$\pm$0.001 & 0.995$\pm$0.001& -0.014$\pm$0.004 & 707760  \\
2.60 $<$R$<$3.20  & 1.011$\pm$0.001 & 1.003$\pm$0.001 & 0.985$\pm$0.001 & 0.998$\pm$0.001& -0.004$\pm$0.002 & 1301720 \\
\hline
 & & & Disc-Dominated subsample & &\\
\hline
1.00 $<$R$<$1.05  & 1.172$\pm$0.021 & 1.056$\pm$0.019   & 0.803$\pm$0.015 & 1.062$\pm$0.019 &  0.151$\pm$0.046 & 4212 \\
1.05 $<$R$<$1.10  & 1.156$\pm$0.026 & 1.040$\pm$0.023   & 0.812$\pm$0.019 & 0.994$\pm$0.022 &  0.013$\pm$0.059 & 2587 \\
1.10 $<$R$<$1.20  & 1.148$\pm$0.017 & 1.035$\pm$0.015   & 0.837$\pm$0.012 & 1.003$\pm$0.015 &  0.018$\pm$0.038 & 6366 \\
1.20 $<$R$<$1.40  & 1.098$\pm$0.010 & 1.036$\pm$0.009   & 0.875$\pm$0.008 & 0.991$\pm$0.009 & -0.029$\pm$0.023 & 17553 \\
1.40 $<$R$<$1.80  & 1.057$\pm$0.005 & 1.021$\pm$0.005   & 0.920$\pm$0.005 & 0.987$\pm$0.005 & -0.028$\pm$0.013 & 56883  \\
1.80 $<$R$<$2.60  & 1.025$\pm$0.002 & 1.012$\pm$0.002   & 0.960$\pm$0.002 & 0.994$\pm$0.003 & -0.013$\pm$0.006 & 228198 \\
2.60 $<$R$<$3.20  & 1.013$\pm$0.002 & 1.006$\pm$0.002   & 0.981$\pm$0.002 & 0.999$\pm$0.002 & -0.002$\pm$0.005 & 370409 \\
\hline

\label{data}

\end{tabular}


\end{table*}

\section{Discussion}

In this paper, we have shown that when haloes are selected in order to contain a
Milky--Way like disc galaxy at their centres the angular momentum of the dark
matter halo is oriented preferentially parallel to the surface of the voids.
Observationally, the same alignment is detected using the baryonic matter
(Trujillo et al. 2006). These two pieces of information are in agreement with
the TTT prediction that both the dark and the baryonic matter component have
conjointly acquired  their angular momentum before the moment of the turnaround.
Interestingly, the signal detected in the real observation c=0.7$^{+0.1}_{-0.2}$
is higher than the one found in the simulations c=0.151$\pm$0.046. This is to be 
expected taking into account that the signal in the dark matter haloes should be
erased by non--linear  effects such as exchange of angular momentum between the
haloes. Future work, consequently, should explore the strength of the alignment
of the haloes at the moment of  turnaround. At that early epoch the strength of
the signal should be as strong as the one measured using the baryonic component.
Porciani et al. (2002b) shows hints that this should be the case by comparing
the relation between the halo spin and the linear shear tensor at different
redshifts from z=50 to z=0.

We have compared our work with previous analysis of the alignment of the
inertia axes and angular momentum with their surrounding large--scale structure
using different simulations. We concentrate first on the alignment of the
angular momentum: using the void framework, neither Heymans et al. (2006) nor
PA06 have found a signal of the alignment of the angular momentum within the
shell of the voids. Their results are easily understood taking into account
that no preselection of the haloes was done in either of these works and that
they explored the signal in just one  shell of width 4 $h^{-1}$ Mpc. In fact,
if we mimic their analysis we find c=-0.014$\pm$0.012 (p=0.996$\pm$0.005), in
agreement with their findings. In the same shell, selecting those haloes with
0$<$B/T$<$0.45 produces c=0.030$\pm$0.021 (p=1.012$\pm$0.008). This is a factor
of 5 weaker than the signal we find in the closest shell to the void surface. 
As we have seen in this work, the signal is only significant at the void
surface, consequently, taking a wide shell can mask it. In the particular case
of Heymans et al. (2006) their mass resolution limit could be another source of
concern, since it is a factor of $\sim$10 worse than the present simulation.

Comparison with other work is more complicated since the analysis of the
alignments is done using a different scheme. BS05 measure the alignment of the
angular momentum along filaments. Interestingly, they found that the angular
momentum of galaxy mass haloes shows a weak tendency  to point along filaments,
while those of group and cluster mass haloes show a very strong tendency to
point perpendicular to the filaments. The significance and strength of their
signal in terms of the c parameter is, however, not quantified. Consequently,
the agreement between ours and their work can be done only qualitatively.
Porciani et al. (2002b) explored the alignment of massive haloes
(M$>$10$^{13}h^{-1}$M$_{\sun}$) between their $final$ spin and the $initial$
shear tensor at the halo position. They found c=0.07$\pm$0.04. The mass regime
explored by these authors and their comparison between an initial and a final
property of the haloes prevents us making a direct comparison between their and
our measurements of the c parameter. Finally, in a recent paper,
Arag\'on--Calvo et al. (2006), using galaxy mass haloes find that the strength
of alignment of their spins in walls is c=0.13$\pm$0.02. This is in excellent
agreement with the value reported in this paper.

If we focus  our attention on the alignment of the inertia axes, we find that
our results are in good agreement with those of BS05, ACC06 and PA06. All these
authors find that the tendency of the minor axis to lie perpendicular to
large--scale filaments is the strongest of the alignments. It is interesting to
note that the strength of these alignments seems to be dependent on the mass of
the haloes, being stronger for the most massive (cluster--sized) ones. BS05 and
ACC06 suggest that the different strength could be related to fact that most
massive haloes receive a larger infall of matter from filaments. This could
also help to explain the tendency of the angular momentum of the most massive
haloes to be perpendicular to the filaments. In this sense, the cluster mass
haloes would acquire most of their current angular momentum from major mergers
along the filaments, whereas the angular momentum of the galaxy mass haloes
will still have memory of the initial tidal fields.

The alignment of the haloes with their local large--scale structure is not only
of interest to constrain models of galaxy formation, it could also be relevant
to explain other observational features. For example, the tendency of
satellite galaxies to avoid orbits that are coplanar with their host spiral
galaxies (know as the ''Holmberg effect'') found in observations (Holmberg
1969; Zaritsky et al. 1997; Sales \& Lambas 2004, but see Agustsson \& Brainerd
2006) and in simulations (Zentner et al. 2005; Libeskind et al. 2006). This
could be due to  the preferential accretion of satellites along filaments, that
we have seen are preferentially aligned with the major axis of the host halo.

Finally, it is worth  pointing out the potential importance of the alignments we
have discussed here to strong and weak lensing studies. In particular,
these alignments could contribute to the weak lensing signal producing a
shear--ellipticity correlation (Hirata \& Seljak 2004). The degree of
contamination that these alignments will produce in the weak lensing surveys
should be explored in future work.

\section{Acknowledgments}

We would like to acknowledge very helpful discussions with J. E. Betancort, C.
Carretero and V. Springel. We thank the anonymous referee for her/his useful
comments. The Millennium simulation was performed at the MPA by the Virgo
Consortium, much of the analysis presented here was completed at the Nottingham
HPC centre.


\begin{thebibliography}{99}

\bibitem[]{} Agustsson I., Brainerd, T. G., 2006, ApJ, 644, L25
\bibitem[]{} Arag\'on-Calvo M. A., van de Weygaert R., Jones B. J. T., van der
Hulst, J. M. T., 2006, ApJ, submitted, astro-ph/0610249 
\bibitem[]{} Avni Y., \& Bahcall J.N., 1980, ApJ, 235, 694
\bibitem[]{} Altay G., Colberg J. M., Croft R.A.C, 2006, MNRAS, 370, 1422 (ACC06)
\bibitem[]{} Bailin J. \& Steinmetz M., 2005, ApJ, 627, 647 (BS05)
\bibitem[]{} Barnes J., Efstathiou G. P., 1987, ApJ, 319, 575
\bibitem[]{} Basilakos S., Plionis M., Yepes G., Gottl\"ober S., Turchaninov V.,
  2006, MNRAS, 365, 539
\bibitem[]{} Bingelli B., 1982, A\&A, 107, 338
\bibitem[]{} Bond J. R., Kofman L., Pogosyan D., 1996, Nature, 380, 603
\bibitem[]{} Catelan P., Theuns T., 1996, MNRAS, 282, 436
\bibitem[]{} Colless M. et al., 2001, MNRAS, 328, 1039
\bibitem[]{} Croft R. A. C., Metzler C. A., 2000, ApJ, 545, 561
\bibitem[]{} Croton D. J. et al., 2006, MNRAS, 365, 11
\bibitem[]{} Debattista V. P., Sellwood J. A., 1999, ApJ, 513, L107
\bibitem[]{} Doroshkevich A. G., 1970, Astrofisika, 6, 581
\bibitem[]{} Fall S.M., Efstathiou G. P., 1980, MNRAS, 193, 189
\bibitem[]{} Faltenbacher A., Gottl\"ober S., Kerscher M., M\"uller V., 2002,
A\&A, 395, 1 
\bibitem[]{} Heavens A. F., Peacock J. A., 1988, MNRAS, 232, 339
\bibitem[]{} Heavens A. F., Refregier A., Heymans C., 2000, MNRAS, 319, 649
\bibitem[]{} Hernquist L., Katz N., 1989, ApJS, 70, 419
\bibitem[]{} Heymans C., White M., Heavens A., Vale, C., Van Waerbeke L., 2006,
MNRAS, 371, 750
\bibitem[]{} Hirata C.M., Seljak U., 2004, Phys. Rev. D, 70
\bibitem[]{} Holmberg E., 1969, Ark. Astron., 5, 305
\bibitem[]{} Hopkins P. F., Bahcall N. A., Bode P., 2005, ApJ, 618, 1
\bibitem[]{} Hoyle F., 1951, in Burgers J. M., van den Hulst H. C., eds,
Problems of Cosmical Aerodynamics, Central Air Documents, Dayton, Ohio, p. 195 
\bibitem[]{} Kasun S. F., Evrard A. E., 2005, ApJ, 629, 781
\bibitem[]{} Lee J., Pen U., 2000, ApJ, 532, L5
\bibitem[]{} Lee J., Pen U., 2002, ApJ, 567, L111
\bibitem[]{} Lee J., 2004, ApJ, 614, L1
\bibitem[]{} Lee J., Kang X., Jing Y. P., 2005, ApJ, 629, L5 
\bibitem[]{} Libeskind N. I., Cole S., Frenk C. S., Okamoto T., Jenkins A.,
2006, MNRAS, submitted, astro-ph/0607237
\bibitem[]{} L\'opez-Corredoira M., Betancort-Rijo J., Beckman J. E., 2002,
A\&A, 386, L169
\bibitem[]{}  Monaghan J.J., Lattanzio J. C., 1985, A\&A, 149, 135
\bibitem[]{} Navarro J.F., Abadi M. G., Steinmetz M., 2004, ApJ, 613, L41
\bibitem[]{} Onuora L. I., Thomas P. A., 2000, MNRAS, 319, 614
\bibitem[]{} Ostriker E. C., Binney J. J., 1989, MNRAS, 237, 785
\bibitem[]{} Patiri S. G., Betancort--Rijo J., Prada F., Klypin
A., Gottl\"ober S., 2006a, MNRAS, 369, 335
\bibitem[]{} Patiri S. G., Cuesta A. J., Prada F., Betancort--Rijo J., Klypin
A., 2006b, ApJ, in press, astro-ph/0606415 (PA06)
\bibitem[]{} Porciani C., Dekel A., Hoffman Y., 2002a, MNRAS, 332, 325 
\bibitem[]{} Porciani C., Dekel A., Hoffman Y., 2002b, MNRAS, 332, 339 
\bibitem[]{} Peebles P.J.E., 1969, ApJ, 155, 393
\bibitem[]{} Sackett P. D., 1999, in ASP Conf. Ser. 182, Galaxy Dynamics, ed. D.
Merritt, J. A. Sellwood, M. Valluri (San Francisco: ASP), 393
\bibitem[]{} Sales L., Lambas D. G., 2004, MNRAS, 348, 1236
\bibitem[]{} Spergel D.N. et al., 2003, ApJS, 148, 175 
\bibitem[]{} Splinter R. J., Melott, A. L., Linn A. M., Buck C., Tinker J., 1997, ApJ, 479, 632
\bibitem[]{} Springel V. et al., 2005, Nature, 435, 629
\bibitem[]{} Sugerman B., Summers F. J., Kamionkowski M., 2000, MNRAS, 311,
762
\bibitem[]{} Thomas P.A. et al., 1998, MNRAS, 296, 1061 
\bibitem[]{} Trujillo I., Carretero C., Patiri S.G., 2006, ApJ, 640, L11
\bibitem[]{} van Haarlem M., van de Weygaert R., 1993, ApJ, 418, 544
\bibitem[]{} White S.D.M., 1984, ApJ, 286, 38
\bibitem[]{} Yang X., van den Bosch F.C., Mo H.J., Mao S., Kang X., Weinmann
S.M., Guo Y., Jing Y.P., 2006, MNRAS, 369, 1293
\bibitem[]{} Zaritsky D., Smith R., Frenk C. S., White S. D. M., 1997, ApJ, 478, L53 
\bibitem[]{} Zentner A.R., Kravtsov A. V., Gnedin, O. Y., Klypin, A. A., 2005,
ApJ, 629, 219 


\end{thebibliography}
\end{document}